\documentstyle[preprint,aps]{revtex}
\tightenlines
\begin{document}
\draft
\title
{Thermodynamics for Fractal Statistics}
\author
{Wellington da Cruz\footnote{E-mail: wdacruz@exatas.campus.uel.br}} 
\address
{Departamento de F\'{\i}sica,\\
 Universidade Estadual de Londrina, Caixa Postal 6001,\\
Cep 86051-990 Londrina, PR, Brazil\\}
\date{\today}
\maketitle
\begin{abstract}
We consider for an anyon gas its thermodynamic
properties taking into account the fractal statistics 
obtained by us recently. This approach describes the 
anyonic excitations in terms of equivalence classes 
labeled by fractal parameter or Hausdorff dimension $h$. 
An exact equation of state is obtained in the high-temperature 
and low-temperature limits, for gases with a constant 
density of states.

\end{abstract}



In the literature, the fractional spin particles are a subject 
of attention by now for a long date. Anyons\cite{R1} are such 
kind of particles, that live in two dimensions, where 
wavefunctions aquire an arbitrary phase when two of them are braided. 
A fractional exclusion statistics was proposed in\cite{R2} 
for arbitrary dimensions, which interpolates between bosonic 
and fermionic statistics. In\cite{R3} was showed that anyons 
also satisfy an exclusion principle when Haldane statistics is 
generalised to infinite dimensional Hilbert spaces. The statistics 
parameter $g$ was introduced  for particles with finite dimensional 
Hilbert spaces and defined as $g=-\frac{\Delta{D}}
{\Delta{N}}$, such that if 
we add $\Delta{N}$ particles into the system, the number 
of single-particles states $D$ available for further 
particles is altered by $\Delta{D}$. This expresses a 
reduction of single-particle Hilbert space by the exclusion 
statistics parameter $g$. In\cite{R4} a distribution function for 
particles obeying a generalised exclusion principle was introduced, 
and the statistical mechanics and thermodynamics considered. 
Also in\cite{R5}, some physical properties considering 
exclusion statistics for identical particles 
called $g$-ons was obtained.

On the other hand, in the context of Fractional 
Quantum Hall Effect (FQHE), the anyonic model can be considered. 
FQHE are observed in two-dimensional eletron 
systems in a stronger magnetic field at low 
temperature, and for such systems in\cite{R6} 
wavefunctions were obtained for the lowest 
Landau level ( LLL). FQHE quasiparticles as anyons confined 
in the LLL were discussed in terms of exclusion 
statistics in\cite{R7}.

Recently, we proposed\cite{R8} another 
way to consider anyons. They can be classified 
in terms of a fractal parameter or Hausdorff 
dimension $h$, and so these particles appear into 
the equivalence classes labeled by $h$. In this way 
{\it ab initio} we consider spin $s$ for such 
particles which are related to statistics $\nu$, 
by $\nu=2s$. The 
particles within the equivalence class $h$ 
satisfy a specific fractal statistics. This constitutes
 in some sense a more suitable approach to 
 fractional spin particles, since unify them naturally. 
 A connection with 
 FQHE systems 
also was made, such that the filling factors for which 
FQHE were 
observed just appear into these classes and the 
ocurrence of FQHE were 
considered in terms of duality between equivalence classes. A 
supersymmetric character between fermions and bosons 
( dual objects by definition ) in the extremes 
of the range of $h$ is a special result 
advanced by this formulation.

In this Letter, we consider the thermodynamics for 
fractal statistics and in this way a free gas of anyons 
are termed fractons $(h,\nu)$. In particular, at low 
temperature and in the low density, for a gas with a 
constant density of states in energy, equation of state for 
such systems are obtained exactly. 

The fractal parameter 
$h$ is defined into the interval 
$1$$\;$$ < $$\;$$h$$\;$$ <$$\;$$ 2$. In the extremes 
we have Fermi and Bose statistics, respectively\cite{R8}. 
The average ocuppation number is given by

\begin{eqnarray}
\label{e.9}
n_{j}=\frac{1}{{\cal{Y}}(\xi_{j})-h},
\end{eqnarray}

\noindent where the function ${\cal{Y}}(\xi_{j})$ 
satisfies the functional 
equation
\begin{eqnarray}
\label{e.67}
\xi_{j}=\left\{{\cal{Y}}(\xi_{j})-1\right\}^{h-1}
\left\{{\cal{Y}}(\xi_{j})-2\right\}^{2-h},
\end{eqnarray}
 
 \noindent with $\xi_{j}=\exp
 \left\{(\epsilon_{j}-\mu)/KT\right\}$.

The fractal parameter $h$ is related 
 to statistics $\nu$, as follows:  

\begin{eqnarray}
\label{e.2}
&&h-1=1-\nu,\;\;\;\; 0 < \nu < 1;\;\;\;\;\;\;\;\;
 h-1=\nu-1,\;
\;\;\;\;\;\; 1 <\nu < 2;\;\nonumber\\
&&h-1=3-\nu,\;\;\;\; 2 < \nu < 3;\;\;\;\;\;\;\;\;
h-1=\nu-3,\;\;\;\;\;\;\; 3 < \nu < 4;\;\nonumber\\
&&h-1=5-\nu,\;\;\;\; 4 < \nu < 5\;;\;\;\;\;\;\;\;
h-1=\nu-5,\;\;\;\;\;\;\; 5 < \nu < 6;\;\\
&&h-1=7-\nu,\;\;\;\;\; 6 < \nu < 7;\;\;\;\;\;\;\;
h-1=\nu-7,\;\;\;\;\;\;\; 7 < \nu < 8;\;\nonumber\\
&&h-1=9-\nu,\;\;\;\;8 < \nu < 9;\;\;\;\;\;\;\;\;
h-1=\nu-9,\;\;\;\;\;\;\; 9 < \nu < 10;\nonumber\\
&&etc,\nonumber
\end{eqnarray}

\noindent and so, we have a mirror symmetry 
for all spectrum of $\nu$. This symmetry 
gives us, for example, the equivalence classes

\begin{eqnarray}
&&\left\{\frac{1}{3},\frac{5}{3},\frac{7}{3},
 \frac{11}{3},\cdots\right\}_
{h=\frac{5}{3}},\;\;\;\;\;\;\;\;\;\;\left\{\frac{2}{3},
\frac{4}{3},\frac{8}{3},\frac{10}{3},\cdots\right\}_
{{\tilde{h}}=\frac{4}{3}};\nonumber\\
&&\left\{\frac{1}{2},\frac{3}{2},\frac{5}{2},
 \frac{7}{2},\cdots\right\}_
{h={\tilde{h}}=\frac{3}{2}} ;\nonumber\\
&&\left\{\frac{1}{5},\frac{9}{5},\frac{11}{5},
 \frac{19}{5},\cdots\right\}_
{h=\frac{9}{5}},\;\;\;\;\;\;\;\;\left\{\frac{4}{5},
\frac{6}{5},\frac{14}{5},\frac{16}{5},\cdots\right\}_
{{\tilde{h}}=\frac{6}{5}};\nonumber\\
&&\left\{\frac{6}{7},\frac{8}{7},\frac{20}{7},
 \frac{22}{7},\cdots\right\}_
{h=\frac{8}{7}},\;\;\;\;\;\;\;\left\{\frac{1}{7},
\frac{13}{7},\frac{15}{7},\frac{27}{7},\cdots\right\}_
{{\tilde{h}}=\frac{13}{7}};\nonumber\\
&&\left\{\frac{2}{7},\frac{12}{7},\frac{16}{7},
 \frac{26}{7},\cdots\right\}_
{h=\frac{12}{7}},\;\;\;\;\;\left\{\frac{5}{7},
\frac{9}{7},\frac{19}{7},\frac{23}{7},\cdots\right\}_
{{\tilde{h}}=\frac{9}{7}};\nonumber\\
&&\left\{\frac{2}{9},\frac{16}{9},\frac{20}{9},
 \frac{34}{9},\cdots\right\}_
{h=\frac{16}{9}},\;\;\;\;\;\left\{\frac{7}{9},
\frac{11}{9},\frac{25}{9},\frac{29}{9},\cdots\right\}_
{{\tilde{h}}=\frac{11}{9}};\nonumber\\
&&\left\{\frac{2}{5},\frac{8}{5},\frac{12}{5},
 \frac{18}{5},\cdots\right\}_
{h=\frac{8}{5}},\;\;\;\;\;\;\;\;\left\{\frac{3}{5},
\frac{7}{5},\frac{13}{5},\frac{17}{5},\cdots\right\}_
{{\tilde{h}}=\frac{7}{5}};\\
&&\left\{\frac{3}{7},\frac{11}{7},\frac{17}{7},
 \frac{25}{7},\cdots\right\}_
{h=\frac{11}{7}},\;\;\;\;\;\left\{\frac{4}{7},
\frac{10}{7},\frac{18}{7},\frac{24}{7},\cdots\right\}_
{{\tilde{h}}=\frac{10}{7}};\nonumber\\
&&\left\{\frac{4}{9},\frac{14}{9},\frac{22}{9},
 \frac{32}{9},\cdots\right\}_
{h=\frac{14}{9}},\;\;\;\;\;\left\{\frac{5}{9},
\frac{13}{9},\frac{23}{9},\frac{31}{9},\cdots\right\}_
{{\tilde{h}}=\frac{13}{9}};\nonumber\\
&&\left\{\frac{6}{13},\frac{20}{13},\frac{32}{13},
 \frac{46}{13},\cdots\right\}_
{h=\frac{20}{13}},\;\;\;\left\{\frac{7}{13},
\frac{19}{13},\frac{33}{13},\frac{45}{13},\cdots\right\}_
{{\tilde{h}}=\frac{19}{13}};\nonumber\\
&&\left\{\frac{5}{11},\frac{17}{11},\frac{27}{11},
 \frac{39}{11},\cdots\right\}_
{h=\frac{17}{11}},\;\;\;\left\{\frac{6}{11},
\frac{16}{11},\frac{28}{11},\frac{38}{11},\cdots\right\}_
{{\tilde{h}}=\frac{16}{11}};\nonumber\\
&&\left\{\frac{7}{15},\frac{23}{15},\frac{37}{15},
 \frac{53}{15},\cdots\right\}_
{h=\frac{23}{15}},\;\;\;\left\{\frac{8}{15},
\frac{22}{15},\frac{38}{15},\frac{52}{15},\cdots\right\}_
{{\tilde{h}}=\frac{22}{15}},\nonumber
\end{eqnarray}

\noindent where $h$ and ${\tilde{h}}$ are dual classes 
defined by

\begin{equation} 
\label{e.10}
\tilde{h}=3-h.
\end{equation}

This is another symmetry that relates 
statistics $(\nu,{\tilde{\nu}})$, 
such that in the context of FQHE this corresponds to 
filling factors ( rational numbers with an 
odd denominator ), that is, 
parameters which characterize the Hall resistance in 
two-dimensional eletron systems in a stronger magnetic 
field at low temperature. In particular, the class 
$h={\tilde{h}}=\frac{3}{2}$ is a selfdual class and 
for particles into this class we obtain the distribution

\begin{equation}
\label{e.23}
n=\frac{1}{\sqrt{\frac{1}{4}+\xi^2}}.
\end{equation}

The thermodynamic properties for a given class $h$ 
can be now obtained. In two dimensions, the particle number in a 
volume element ${d^2}p\;dV$ in phase space is given by

\begin{eqnarray}
\label{e.11}
dN={\cal{G}}\frac{d^2p\;dV}{(2\pi {\hbar})^2}
\frac{1}{{\cal{Y}}(\xi)-h},
\end{eqnarray}

\noindent with ${\cal{G}}=\nu+1$ and $\nu=2s$. Integrating over
 $V(area)$, we obtain

\begin{eqnarray}
\label{e.11}
dN_{p}=\frac{{\cal{G}}\;V}{2\pi {\hbar}^2}
\frac{p\;dp}{{\cal{Y}}(\xi)-h},
\end{eqnarray}

and for the dispersion $\epsilon(p)=\frac{p^2}{2m}$

\begin{eqnarray}
\label{e.12}
dN_{\epsilon}=\frac{m\;{\cal{G}}\;V}{2\pi{\hbar}^2}
\frac{d\epsilon}{{\cal{Y}}[\xi(\epsilon)]-h}.
\end{eqnarray}

The total energy is given by

\begin{eqnarray}
\label{e.13}
E&=&\int_{0}^{\infty}\epsilon\; dN_{\epsilon}\nonumber\\
&=&\frac{m\;{\cal{G}}\;V}{2\pi{\hbar}^2}\int_{0}^{\infty}
\frac{\epsilon\; d\epsilon}{{\cal{Y}}[\xi(\epsilon)]-h}
\end{eqnarray}

\noindent and  

\begin{equation}
\label{e.13}
PV=E=\frac{m\;{\cal{G}}\;V}{2\pi{\hbar}^2}\int_{0}^{\infty}
\frac{\epsilon\; d\epsilon}{{\cal{Y}}[\xi(\epsilon)]-h}.
\end{equation}

\noindent For the thermodynamic potential we obtain

\begin{equation}
\label{e.13}
\Omega=-PV=-\frac{m\;{\cal{G}}\;V}{2\pi{\hbar}^2}\int_{0}^{\infty}
\frac{\epsilon\;d\epsilon}{{\cal{Y}}[\xi(\epsilon)]-h}.
\end{equation}

\noindent For the class $h=\frac{3}{2}$ and 
$\nu=\frac{1}{2}$, the pressure is given by

\begin{equation}
\label{e.13}
P=\frac{3}{2}\frac{m}{2\pi{\hbar}^2}\int_{0}^{\infty}
\frac{\epsilon\;d\epsilon}
{\sqrt{\frac{1}{4}+\exp{2(\epsilon-\mu)/KT}}},
\end{equation}

\noindent and defining the variable $z=\frac{\epsilon}{KT}$, 
we write down 

\begin{equation}
\label{e.14}
\frac{P}{KT}=\frac{3}{2}\frac{m\;K\;T}{2\pi{\hbar}^2}\int_{0}^{\infty}
\frac{2\;z\;dz}
{\sqrt{1+4\exp{2(z-\mu/KT)}}},
\end{equation}

\noindent which combined with the particle density

\begin{equation}
\label{e.15}
\frac{N}{V}=\frac{3}{2}\frac{m\;K\;T}{2\pi{\hbar}^2}\int_{0}^{\infty}
\frac{2\;dz}
{\sqrt{1+4\exp{2(z-\mu/KT)}}},
\end{equation}

\noindent determines the relation between $P,V$ and $T$, that is, the
 equation of state for a gas of free particles termed {\it fracton} $(h,\nu)
 =\left(\frac{3}{2},\frac{1}{2}\right)$.
 
 From the Eq.(\ref{e.67}) we see that the function ${\cal{Y}}(\xi)$ get 
 its degree from $h$ denominator and so for calculate the roots of 
 ${\cal{Y}}$ in terms of the parameter $\xi$ can be a 
 formidable mathematical task. Otherwise, we can solve ${\cal{Y}}(\xi)$ 
 numerically. For example, the classes $h=\frac{4}{3},\frac{5}{3}$ have 
 a third degree algebraic equation
 
\begin{equation}
\label{e.16}
{{\cal{Y}}}^3+a_{1}{{\cal{Y}}}^2+a_{2}{\cal{Y}}+a_{3}=0,
\end{equation}

\noindent which has real solution

\begin{equation}
\label{e.17}
{\cal{Y}}(\xi)=s+t-\frac{a_{1}}{3},
\end{equation}

\noindent where

\begin{eqnarray}
\label{e.18}
&&s=\left\{{\sqrt{r+{\sqrt{q^3+r^2}}}}\right\}^{3},\nonumber\\
&&t=\left\{{\sqrt{r-{\sqrt{q^3+r^2}}}}\right\}^{3}\nonumber
\end{eqnarray}

\noindent and

\begin{eqnarray}
&&q=\frac{3a_{2}-{a_{1}}^2}{9},\nonumber\\
&&r=\frac{9a_{1}a_{2}-27a_{3}-2{a_{1}}^3}{54}.\nonumber
\end{eqnarray}

\noindent For the class $h=\frac{4}{3}$ we obtain

\begin{equation}
{{\cal{Y}}}^3-5{{\cal{Y}}}^2+8{\cal{Y}}-(4+\xi^{3})=0,
\end{equation}

\noindent and for $h=\frac{5}{3}$

\begin{equation}
{{\cal{Y}}}^3-4{{\cal{Y}}}^2+5{\cal{Y}}-(2+\xi^{3})=0.
\end{equation}

\noindent The average ocuppation number is given by

\begin{equation}
\label{e.29}
n=\frac{1}{s+t-\frac{1}{3}a_{1}-h}
\end{equation}

\noindent and the thermodynamic properties for the 
{\it fractons} $(h,\nu)=\left(\frac{4}{3},\frac{2}{3}\right);
\left(\frac{5}{3},\frac{1}{3}
\right)$, can be considered. However, 
we can follow another way. The 
distribution function can be written 
in terms of the single-state grand 
partition function ${\Theta}_{j}$, as

\begin{equation}
n_{j}=\xi_{j}\frac{\partial\ln\Theta_{j}}{\partial{{\xi}_{j}}},
\end{equation}

\noindent where $\Theta_{j}=\frac{{\cal{Y}}_{j}-2}{{\cal{Y}}_{j}-1}$. 
On the other hand, we can expand $\Theta_{j}$ in powers of $\xi_{j}$,

\begin{eqnarray}
\Theta_{j}&=&\frac{{\cal{Y}}_{j}-2}{{\cal{Y}}_{j}-1}\nonumber\\
&=&\sum_{l=0}^{\infty}V_{l}\;\xi_{j}^{l}
\end{eqnarray} 

\noindent and

\begin{eqnarray}
n_{j}=\sum_{l=1}^{\infty}U_{l}\;\xi_{j}^{l},
\end{eqnarray}

\noindent where

\begin{eqnarray}
V_{l}=\prod_{k=2}^l\left\{1+(h-2)\frac{l}{k}\right\};\;
U_{l}=\prod_{k=1}^{l-1}\left\{1+(h-2)\frac{l}{k}\right\}.\nonumber
\end{eqnarray}

\noindent These coefficients are obtained 
in the same way as discussed in\cite{R9}. They do not depend on 
$G$ states number and are calculated from the 
combinatorial formula  
 
\begin{equation}
\label{e.23}
w=\frac{\left[G+(h-1)N-1\right]!}{N!
\left[G+(h-2)N\right]!},
\end{equation}

\noindent taking into account the grand partition function
 
\begin{eqnarray}
\label{e.1}
{\cal{Z}}(G,\xi)&=&\sum_{N=0}^{\infty}w(G,N)\;\xi^N\\
&=&\left(\sum_{l}V_{l}\;\xi^l\right)^G\nonumber,
\end{eqnarray}

\noindent that is as $G$-th power of the 
single-state partition function $\Theta$. 

The Eq.(\ref{e.23}) 
gives the same statistical mechanics as 
that introduced in\cite{R8}

\begin{equation}
\label{e.100}
w^{\prime}=\frac{\left[G+(N-1)(h-1)\right]!}{N!
\left[G+(N-1)(h-1)-N\right]!}.
\end{equation}

Now, we can consider at low temperature Sommerfeld 
expansions\cite{R10}. In this way, we handle 
integrals of the type

\begin{eqnarray}
{\cal{J}}[f]=KT\int_{2}^{1}\frac{d{\cal{Y}}}{\left({\cal{Y}}
-1\right)\left({\cal{Y}}-2\right)}
\;f\left\{\mu+KT
\left[h\ln\left\{\frac{{\cal{Y}}-1}{{\cal{Y}}-2}\right\}
-\ln\left\{\frac{{\cal{Y}}-1}{\left({\cal{Y}}-2\right)^2}
\right\}\right]\right\},
\end{eqnarray}

\noindent and after expansion we have

\begin{eqnarray}
{\cal{J}}[f]=h^{-1}\int_{0}^{\mu}\;f(\epsilon)
\;d\epsilon+\sum_{l=0}^{\infty}
\frac{(KT)^{l+1}}{l!}C_{l}(h)f^{(l)}(\mu),
\end{eqnarray}

\begin{eqnarray}
C_{l}(h)=\sum_{k=0}^{l-1}C_{l,k}\;h^k,
\end{eqnarray}

\noindent with

\begin{eqnarray}
C_{l,k}=(-)^{l-k}
\left(\begin{array}{c}
l\\k
\end{array}\right)
\int_{2}^{1}\frac{d{\cal{Y}}}{\left({\cal{Y}}-1\right)
\left({\cal{Y}}-2\right)}
\ln^{l-k}\left\{\frac{{\cal{Y}}-1}{{\cal{Y}}-2}\right\}\ln^{k}
\left\{\frac{{\cal{Y}}-1}
{({\cal{Y}}-2)^2}\right\}.
\end{eqnarray}

We can obtain now some thermodynamic quantities. 
Let us consider, 
the dispersion as $\epsilon(p)=ap^\sigma$, such that 
in D-dimension we get for a few 
terms of $C_{l}$, the expressions

\begin{eqnarray}
\frac{\cal{E}}{\gamma\;V}=\frac{\mu^{\frac{D}{\sigma}+1}}
{h\left(\frac{D}{\sigma}+1\right)}+
KTC_{0}(h)\mu^{\frac{D}{\sigma}}+\frac{1}
{2}(KT)^2C_{1}(h)
\left(\frac{D}{\sigma}\right)\mu^{\frac{D}{\sigma}-1}+\cdots,
\end{eqnarray}

\begin{eqnarray}
\frac{{\cal{N}}}{\gamma\;V}=\frac{\mu^{\frac{D}{\sigma}}}
{h\left(\frac{D}{\sigma}\right)}+
KTC_{0}(h)\mu^{\frac{D}{\sigma}-1}+\frac{1}
{2}(KT)^2C_{1}(h)
\left(\frac{D}{\sigma}-1\right)\mu^{\frac{D}{\sigma}-2}+\cdots,
\end{eqnarray}

\noindent for the energy and the 
particle number, where $\gamma=\frac{m\;{\cal{G}}}{2\pi\hbar^2}$,
 for $D=\sigma=2$. The coefficients 
 $C_{0}(h)$,$C_{1}(h)$ are given by

\begin{eqnarray}
C_{0}(h)&=&C_{0,0}=\int_{2}^1\frac{d{\cal{Y}}}
{({\cal{Y}}-1)({\cal{Y}}-2)};\\
C_{1}(h)&=&C_{1,0}=-\int_{2}^1\frac{d{\cal{Y}}}
{({\cal{Y}}-1)({\cal{Y}}-2)}
\ln\left\{\frac{{\cal{Y}}-1}{{\cal{Y}}-2}\right\}.\nonumber
\end{eqnarray}

\noindent We can show that to second order $T$

\begin{eqnarray}
&&{\cal{E}}={\cal{E}}_{0}\left\{1+h\;\frac{KT}{\mu_{0}}C_{0}(h)+
2h\;\frac{(KT)^2}{\mu_{0}^2}C_{1}(h)\left(\frac{D}{\sigma}\right)
-h\;C_{0}^2(h)\frac{(KT)^2}{\mu_{0}^2}
\left(\frac{D}{\sigma}\right)
\left(\frac{D}{\sigma}+1\right)\right\},\\
&&\mu=\mu_{0}\left\{1-h\;\frac{KT}{\mu_{0}}
\left(\frac{D}{\sigma}\right)C_{0}(h)
-h\;\frac{(KT)^2}{\mu_{0}^2}C_{1}(h)
\left(\frac{D}{\sigma}\right)\left(\frac{D}
{\sigma}-1\right)\right\},
\end{eqnarray}

\noindent for the energy and chemical potential, 
respectively; and 
to order $T$

\begin{eqnarray}
&&{\cal{C}}(h)=h\;{\cal{E}}_{0}\frac{K}{\mu_{0}}C_{0}(h)
+2h\;{\cal{E}}_{0}\frac{K^2T}{\mu_{0}^2}C_{1}(h)\left(
\frac{D}{\sigma}\right)-h\;
{\cal{E}}_{0}\frac{K^2T}{\mu_{0}^2}C_{0}^2(h)
\left(\frac{D}{\sigma}\right)
\left(\frac{D}{\sigma}+1\right),
\end{eqnarray}

\noindent for the specific heat, where 
$\mu_{0}$ and ${\cal{E}}_{0}$ are the 
zero-temperature chemical potential and energy.

For a dilute gas or high-temperature regime, we can 
investigate the thermodynamic properties of a 
free gas using virial expansion, that is the 
equation of state relates $P$, $T$ and 
$\rho$ ( pressure, temperature 
and density respectively ) as follows\cite{R11}

\begin{equation}
P=\rho\;K\;T\left\{1+a_{2}\rho+a_{2}\rho^2+\cdots\right\},
\end{equation}

\noindent where $a_{2}$,$a_{3}$, etc are 
the virial cofficients. On the other hand, the 
grand potential is the sum over the single-particle states

\begin{equation}
\Omega=-KT\sum_{j}\ln\Theta_{j},
\end{equation}

\noindent and as $\Omega=-PV$, we have

\begin{equation}
\frac{P}{KT}=\sum_{j}\frac{1}{V}\ln\Theta_{j},
\end{equation}

\noindent and expanding in terms of $\xi_{j}$ 
after summation over $j$, we get the cluster expansion

\begin{equation}
P=KT\sum_{l=1}^{\infty}b_{l}\;z^{l},
\end{equation}

\noindent and for density

\begin{equation}
\rho=\sum_{l=1}^{\infty}l\;b_{l}\;z^{l},
\end{equation}

\noindent where $b_{l}=\prod_{k=1}^{l-1}
\left\{1+(h-2)\frac{l}{k}\right\}
\frac{{\cal{Z}}_{1}(\frac{T}{l})}{l}$,  
$z=\exp\frac{\mu}{KT}$ is the fugacity and 
${\cal{Z}}_{1}(T)=V\gamma(KT)^{\frac{D}{\sigma}}$ 
is the one-particle partition function $\sum_{j}
\exp\left\{-\frac{\epsilon_{j}}{KT}\right\}$. In this 
way we can show that

\begin{equation}
a_{2}=2^{-\frac{D}{\sigma}}\left(h-\frac{3}{2}\right)\frac{V}
{{\cal{Z}}_{1}(T)},
\end{equation}

\begin{equation}
a_{3}=\left\{\left(5-3h\right)\left(3h-4\right)
3^{-(\frac{D}{\sigma}+1)}+
2^{-2(\frac{D}{\sigma})}\left(2h-3\right)^2 
\right\}\left[\frac{V}
{{\cal{Z}}_{1}(T)}\right]^2,
\end{equation}

\noindent with $a_{l}={\tilde{a}}_{l}V^{l-1}$,$\;$ 
${\tilde{b}}_{l}=\frac{b_{l}}{b_{1}^l}$,$\;$ 
${\tilde{a}}_{2}=-{\tilde{b}}_{2}$,$\;$ 
${\tilde{a}}_{3}=-2{\tilde{b}}_{3}+4{\tilde{b}}_{2}^2$, etc.

For a gas with a constant density of states in energy, 
with $D=\sigma=2$, we have

\begin{equation}
\frac{{\cal{Y}}(0)-1}{{\cal{Y}}(0)-2}=e^{-\frac{\mu}{hKT}}=
e^{-\frac{\rho}{\gamma\;KT}}
\end{equation}

\noindent and from the Eq.(\ref{e.67}) we obtain

\begin{equation}
\mu(\rho,T)=\frac{(h-1)\rho}{\gamma}+KT\ln\left\{e^{-\frac{\rho}
{\gamma\;KT}}-1\right\}
\end{equation}

\noindent and in the low density, the pressure is a finite expression

\begin{equation}
P=\left(h+1\right)\frac{\rho^2}
{2\gamma}
\end{equation}

\noindent and at low temperature, we obtain

\begin{equation}
P=\frac{h\rho^2}{2\gamma}+\gamma(KT)^2C_{1}(h).
\end{equation}

In summary, we have considered a free gas 
of fractons $(h,\nu)$, that is anyons classified 
into equivalence classes labeled by fractal parameter $h$. 
Fractons satisfy a fractal statistics Eq.(\ref{e.9}). 
The thermodynamic quantities in 
the low density and at low temperature were considered for 
a gas with a constant density of 
states $ (D=\sigma=2)$ in energy. We have obtained an 
exact equation of state for both regimes. In particular, 
for a gas in the low density the equation of state shows us 
only interaction between pairs of group of two clusters 
of fractons $(h,\nu)$.


\begin{thebibliography}{99}
\bibitem{R1} J. Leinaas and J. Myrheim, Nuovo Cimento {\bf 37}, 
1 (1977); F. Wilczek, Phys. Rev. Lett. {\bf 48}, 1144; 
{\bf 49}, 957 (1982); {\it Fractional Statistics 
and Anyon Superconductivity}, ed. by F. Wilczek 
( World Scientific, 1990)
\bibitem{R2} F. D. M. Haldane, Phys. Rev. Lett. {\bf 67}, 937 
(1991).
\bibitem{R3} M. V. N. Murthy and R. Shankar, 
Phys. Rev. Lett. {\bf 94}, 3331 
(1994).
\bibitem{R4} Y. S. Wu, Phys. Rev. Lett. {\bf 73}, 922 (1994);
S. B. Isakov, Mod. Phys. Lett. {\bf B8}, 319 (1994); 
A. K. Rajagopal, Phys. Rev. Lett. {\bf 74}, 1048 (1995).
\bibitem{R5} C. Nayak and F. Wilczek, 
Phys. Rev. Lett. {\bf 73}, 2740 (1994).
\bibitem{R6} R. Laughlin, Phys. Rev. Lett. {\bf 50}, 1395 (1983).
\bibitem{R7} see Wu in\cite{R4}; see\cite{R3}; A. 
D. de Veigy and S. Ouvry, Mod. Phys. Lett. 
{\bf B9}, 271 (1995); G. S. Canright and M. D. Johnson, 
J. Phys. {\bf A27}, 3579 (1994); M. V. N. Murthy and R. Shankar, 
Phys. Rev. Lett. {\bf 72}, 3629 (1994).
 \bibitem{R8} W. da Cruz, preprint/UEL-DF/981003, cond-mat/9810381. 
\bibitem{R9} A. P. Polychronakos, Phys. Lett. {\bf B365}, 202 (1996); 
S. B. Isakov, Phys. Rev. {\bf B53}, 6585 (1996); S. B. Isakov 
and S. Mashkevich, Nucl. Phys. 
{\bf B504}, 701 (1997) and references therein.
\bibitem{R10} S. B. Isakov, D. P. Arovas, J. Myrheim and 
A. P. Polychronakos, Phys. Lett. {\bf A212}, 299 (1996); 
see\cite{R5}; E. M. Lifshitz and L. P. Pitaevskii, 
{\it Statistical Physics, Part 1, 
3rd. edition } ( Pergamon Press, Oxford, 1980 ).
\bibitem{R11} K. Huang, {\it Statistical Mechanics} ( John 
Wiley, N. York 1987 )
\end{thebibliography}
\end{document}